\documentclass[
aps,prd,
12pt,
nopreprintnumbers,
showpacs,
eqsecnum,
nofootinbib
]{revtex4-1}

\usepackage{graphicx}
\usepackage{amssymb}

\begin{document}

\title{Spinorial Wheeler--DeWitt wave functions inside black hole horizons}
\author{Nahomi Kan}\email[]{kan@gifu-nct.ac.jp}
\affiliation{National Institute of Technology, Gifu College,
Motosu-shi, Gifu 501-0495, Japan}
\author{Takuma Aoyama}\email[]{b014vbv@yamaguchi-u.ac.jp}
\affiliation{
Graduate School of Sciences and Technology for Innovation, Yamaguchi
University, Yamaguchi-shi, Yamaguchi 753--8512, Japan}
\author{Kiyoshi Shiraishi\footnote{Author to whom any
correspondence should be addressed.}}\email[]{shiraish@yamaguchi-u.ac.jp}
\affiliation{
Graduate School of Sciences and Technology for Innovation, Yamaguchi
University, Yamaguchi-shi, Yamaguchi 753--8512, Japan}
\date{\today}

\begin{abstract}
We revisit the solutions of the Wheeler--DeWitt (WDW) equation inside the
horizons of spherical black holes and planar topological black holes in
arbitrary dimensions. For these systems, the solutions of the equations are found
to have the same form. Therefore, Yeom's \textit{Annihilation-to-nothing}
interpretation can be applied to each case.
We have introduced the Dirac-type WDW equations into quantum cosmology in a recent
paper, so we also apply our formulation to the quantum theory of the interior of
the black hole in order to obtain the solution of the spinorial wave function.
The shape of the wave packet of the spinorial WDW wave function indicates
that the variation of Yeom's interpretation holds in
this scheme.
\end{abstract}



\maketitle

\section{Introduction}
\label{introduction}
Theoretical study on the black hole singularity problem continues with various
approaches. Although the quantum theory of gravity \cite{Kiefer0,Kiefer1} itself
has not been completed,  many authors apply the quantum cosmological approach
(see, for example, Refs.~\cite{CAF,Modesto,AB,CS} and references therein), which
is a method of conducting quantum nature of gravity. This is because inside a
symmetric horizon, the roles of the time ($t$) and the radial coordinates ($r$)
are exchanged, so that the spacetime of the interior of a black hole is
regarded as a time-dependent universe and can be described by the
Kantowski--Sachs metric
\cite{KS}, for instance.

Recently, many works using the minisuperspace Wheeler--DeWitt (WDW) equation
inside the horizon have appeared
\cite{BBCCY,Yeom1,Yeom2,BCY,Perry1,Perry2,Hartnoll}. In particular, Yeom
and his collaborators \cite{BBCCY,Yeom1,Yeom2,BCY} have constructed a concrete
wave packet and proposed the
\textit{Annihilation-to-nothing} interpretation based on its typical behavior. One
of the motivations for the present study is to consider further various types
of wave packets and to reveal the properties of the quantum spacetime inside the
horizon.

In our previous paper, the authors of the present paper introduced the
Dirac wave equation into quantum cosmology on the basis of the extended
minisuperspace \cite{KAHS1,KAHS2}. We now apply our formulation to the quantum
theory inside the horizon. Since this spinorial formulation leads to a positive
definite probability density, it can be applied to further consideration of the
singularity problem.

The rest of the paper is organized as follows.
In Sec.~\ref{sec2}, we review the WDW equations inside the horizon of spherically
symmetric black holes and planar topological black holes in general dimensions.
We obtain their fundamental solutions and construct wave packets using them.
In Sec.~\ref{sec3}, we introduce the Dirac-type wave equation as an alternative to
the WDW equation. We construct the wave packets and compare them with the wave
packets of scalar wave functions obtained in the previous section.
We conclude the present paper with a summary of results and a discussion in the
last section.

\section{A review of the WDW equation inside a black hole}
\label{sec2}
In this section, we review the derivation of the minisuperspace WDW equation
inside a black hole.
First, we consider the Einstein--Hilbert action in $D$-dimensional $(D\ge 4)$
spacetime \cite{Kiefer0,Kiefer1},
\begin{equation}
S=\frac{1}{D-2}\int_\mathcal{M} d^Dx \sqrt{-g}\, R-\frac{2}{D-2}
\int_\mathcal{\partial M}d^{D-1}x\sqrt{h}K\,,
\end{equation}
where $g$ is the determinant of the metric tensor $g_{\mu\nu}$, and
$R$ denotes the scalar curvature constructed from $g_{\mu\nu}$
$(\mu,\nu=0,1,\ldots,D-1)$. The integration in the first term is performed
over the $D$-dimensional spacetime manifold $\mathcal{M}$, while the second
integral is defined on its $(D-1)$-dimensional boundary $\mathcal{\partial M}$.
This second term is known as the Gibbons--Hawking--York boundary term
\cite{York,GH}, which contains the determinant of the $(D-1)$-dimensional metric
on the boundary $h$ and the trace of the extrinsic curvature $K$.

As the metric of the black-hole interior, we assume the Kantowski--Sachs-type
metric \cite{KS}, which represents a homogeneous space with spatial section of
topology
$\approx S^{D-2}\times\mathbf{R}$,
\begin{equation}
ds^2=g_{\mu\nu}dx^\mu
dx^\nu=-e^{2n(t)}dt^2+e^{2\alpha(t)}dr^2+r_s^2e^{2\beta(t)}d\Omega_{D-2}^2\,,
\label{KS}
\end{equation}
where $d\Omega_{D-2}^2$ is the line element of the $(D-2)$-sphere with unit
radius, and $r_s$ is a constant with dimension of length.
The metric (\ref{KS}) is 
related to the Schwarzschild--Tangherlini interior metric \cite{Tangherlini} by,
for example,
\begin{equation}
e^{2\beta}=\frac{t^2}{r_s^2}\,,\quad
e^{2\alpha}=\frac{r_s^{D-3}}{t^{D-3}}-1\,,\quad
e^{2n}=\left(\frac{r_s^{D-3}}{t^{D-3}}-1\right)^{-1}\,,
\end{equation}
where we should notice that the time and radial coordinates are mutually
exchanged from the definition on the exterior solution.

Now, we derive the WDW equation for the interior of a spherical
black hole. We should note that the lapse function $n(t)$ can be arbitrarily
chosen because of the reparametrization invariance on $t$. Thus, we here choose
\begin{equation}
n(t)=\alpha(t)+(D-2)\beta(t)\,.
\label{gf}
\end{equation}
Substituting the metric (\ref{KS}) with the gauge choice (\ref{gf}), we find%
\footnote{The overall constant factor can be
absorbed in the redefinition of the time.}
\begin{equation}
S\propto\int dt \left[-2\dot{\alpha}\dot{\beta}
-(D-3)\dot{\beta}^2+(D-3)r_s^{2(D-3)}e^{2\alpha+2(D-3)\beta}\right]\,.
\end{equation}
Therefore, we obtain the reduced action
\begin{equation}
S\propto \int dt \left[\frac{1}{D-3}
(\dot{X}^2-\dot{Y}^2)+(D-3)r_s^{2(D-3)}e^{2Y}\right]\,,
\end{equation}
where
$X(t)=\alpha(t)$ and $Y(t)=\alpha(t)+(D-3)\beta(t)$,
and consequently we obtain the following Hamiltonian of the system:
\begin{equation}
H(X,Y,\Pi_X,\Pi_Y)=(D-3)\left[\frac{1}{4}(\Pi_X^2-\Pi_Y^2)-r_s^{2(D-3)}e^{2Y}\right]\,,
\end{equation}
where $\Pi_X$ and $\Pi_Y$ are the canonical momenta conjugate to $X$ and $Y$,
respectively.

Now, the Hamiltonian constraint leads to the (minisuperspace) WDW equation
for the scalar wave function $\Phi(X,Y)$,
\begin{equation}
\left(\frac{\partial^2}{\partial X^2}-
\frac{\partial^2}{\partial Y^2}+4r_s^{2(D-3)}e^{2Y}\right)\Phi(X,Y)=0\,,
\label{wdw}
\end{equation}
is obtained after the replacement of $\Pi_X\rightarrow
\hat{\Pi}_X=-i\frac{\partial}{\partial X}$ and
$\Pi_Y\rightarrow \hat{\Pi}_Y=-i\frac{\partial}{\partial Y}$ in the quantum
Hamiltonian constraint $\hat{H}\Phi=0$ \cite{Kiefer0,Kiefer1,Halliwell,Kiefer3}.
Then, we just reproduced the equation presented by Yeom et
al.~\cite{BBCCY,Yeom1,Yeom2,BCY}. Note that this equation is independent of the
spacetime dimensions up to the power of $r_s$.

Before discussing the solutions of the WDW equation,
we consider another system, the interior of a planar topological black hole
\cite{Lemos,HL,Vanzo,Mann,Birmingham}. The action taken here has a negative
cosmological term added,
\begin{equation}
S_\lambda=\frac{1}{D-2}\int d^Dx \sqrt{-g}\, \Bigl[R+2\lambda\Bigr]\,.
\label{nc}
\end{equation}
The metric of the interior of a planar topological black hole
reads
\begin{equation}
ds^2=-\left(\frac{r_s^{D-3}}{t^{D-3}}-\frac{t^2}{L^2}\right)^{-1}dt^2
+\left(\frac{r_s^{D-3}}{t^{D-3}}-\frac{t^2}{L^2}\right)dr^2
+t^2d\bar{\Omega}^2_{D-2}\,,
\end{equation}
which is the classical solution of the Einstein equation obtained from varying
the action (\ref{nc}). Here, the length scale $L$ is given by
$L\equiv\left(\frac{2\lambda}{(D-1)(D-2)}\right)^{-1/2}$, and 
$d\bar{\Omega}^2_{D-2}$ denotes the line element of a flat $(D-2)$ dimensional
torus.

Substituting the previous metric (\ref{KS}), but with $d{\Omega}^2_{D-2}$ 
replaced by
$d\bar{\Omega}^2_{D-2}$, into the action (\ref{nc}), we obtain
\begin{equation}
S\propto \int dt \left[-2\dot{\alpha}\dot{\beta}
-(D-3)\dot{\beta}^2+\frac{D-1}{L^2}r_s^{2(D-2)}e^{2\alpha+2(D-2)\beta}\right]\,.
\end{equation}
Further setting
$X(t)=\alpha(t)-\beta(t)$ and $Y(t)=\alpha(t)+(D-2)\beta(t)$
gives
\begin{equation}
S\propto \int dt \left[\frac{1}{D-1}
(\dot{X}^2-\dot{Y}^2)+\frac{D-1}{L^2}r_s^{2(D-2)}e^{2Y}\right]\,.
\end{equation}
Therefore, the Hamiltonian for this system is found to be
\begin{equation}
H=(D-1)\left[\frac{1}{4}(\Pi_X^2-\Pi_Y^2)-\frac{1}{L^2}r_s^{2(D-2)}e^{2Y}\right]\,,
\end{equation}
and the corresponding WDW equation, which comes from the Hamiltonian constraint,
reads
\cite{Hartnoll}
\begin{equation}
\left(\frac{\partial^2}{\partial X^2}-
\frac{\partial^2}{\partial
Y^2}+\frac{4}{L^2}r_s^{2(D-2)}e^{2Y}\right)\Phi(X,Y)=0\,.
\label{ncwdw}
\end{equation}
Note that the translation $Y-\ln (L/r_s)\rightarrow Y$ just turns the equation
(\ref{ncwdw}) into the equation (\ref{wdw}).%
\footnote{At the classical level, the redefinition of time can also absorb the
constant factor in front of the potential term.}
Thus, we have only to study the WDW equation (\ref{wdw}) for both systems,
the interior of a spherical black hole and that of a planar topological black
hole.

Note that the event horizon is located at $X, Y\rightarrow -\infty$,
while the singularity is located at $X\rightarrow\infty$
and $Y\rightarrow -\infty$, in both cases.

The fundamental solution of (\ref{wdw}) is \cite{BBCCY,Yeom1,Yeom2,BCY}
\begin{equation}
\phi_k(X,Y)= e^{-ikX}K_{ik}(2r_se^Y)\,,
\end{equation}
and general solutions are written by
\begin{equation}
\Phi(X,Y)=\int_{-\infty}^\infty f(k) \phi_k(X,Y)\, dk\,,
\end{equation}
where the function $f(k)$ denotes the amplitude.

Asymptotics of the modified Bessel function of the second
kind with complex order are known to be \cite{Bateman}
\begin{equation}
K_{ik}(z)\sim\sqrt{2\pi}e^{-\frac{k\pi}{2}}
(k^2-z^2)^{-\frac{1}{4}}
\sin\left(\frac{\pi}{4}-\sqrt{k^2-z^2}+k
\cosh^{-1}\frac{k}{z}\right)\,,
\label{asym}
\end{equation}
for $k\rightarrow\infty$.
According to the analysis of Kiefer \cite{Kiefer2}, the peak of the wave
packet tracing the classical path%
\footnote{Of cource, the general solution is written by
$X\rightarrow X+X_0$, where
$X_0$ is a constant, due to the translational invariance of the system.}
 \cite{BBCCY,Yeom1,Yeom2,BCY}
\begin{equation}
e^Y\cosh X=\mbox{const.}\,,
\label{cp}
\end{equation}
can be constructed with the Gaussian amplitude, whose central value in $k$
takes a relatively large value, owing to the asymptotic form (\ref{asym}).

\begin{figure}[ht]
\centering
\includegraphics
{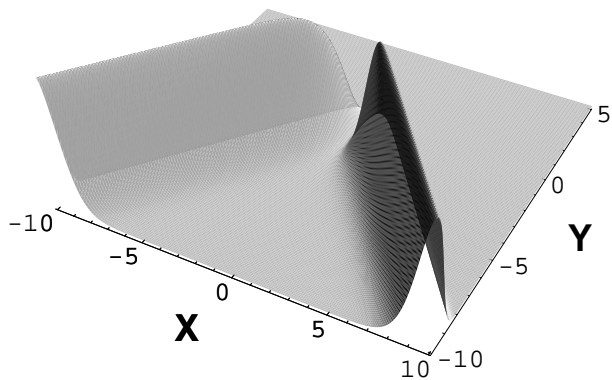}
\caption{Absolute square $|\Phi_1|^2$ of the scalar wave function
for $r_s=1$.
(If you see jaggies in the plot, it is due to resolution limitations.)}
\label{fig1}
\end{figure}

Nevertheless, we can arbitrarily choose the amplitude $f(k)$, so as to reflect
boundary conditions. The Gaussian or rectangular amplitudes are the other
candidates.%
\footnote{Graphical representation of wave packets of similar wave functions
in quantum cosmology are dealt with Refs.~\cite{CGS,ALNW,KKST} et al.}
 Another type of the wave packet with an analytical form is considered
by Refs.~\cite{BBCCY,Yeom1,Yeom2,BCY}, where the simple choice $f(k)=ik$ leads to
\begin{equation}
\Phi_{1}(X,Y)=\int_{-\infty}^\infty k\sin kX\, K_{ik}(2r_s^{D-3}e^Y)\, dk
=2\pi r_s^{D-3}e^Y\sinh X e^{-2r_s^{D-3}e^Y\cosh X}\,.
\end{equation}
The peak of the absolute square of the wave packet $\Phi_1(X,Y)$,
which can be regarded as the probability density
\cite{Kiefer0,Kiefer1,Halliwell,Kiefer3}, exhibits the classical behavior
(\ref{cp}) at the boundaries
$X\rightarrow\pm\infty$, as shown in Fig.~\ref{fig1}.

It is worth observing that the quantum effect seems significant in the vicinity
of $X\approx 0$, where classically there is neither a singularity nor a horizon:
The probability density vanishes here $|\Phi_1(0,0)|^2\approx 0$.
One can interpret this wave packet as an annihilation process of the
trajectory from the horizon ($X=-\infty$) and that from the singularity
($X=\infty$). This interpretation is the so-called
\textit{Annihilation-to-nothing} interpretation proposed by Yeom et
al.~\cite{BBCCY,Yeom1,Yeom2,BCY}.
They interpreted that there are two arrows of time and the two pieces of classical
spacetime are annihilated at a hypersurface inside the horizon.

In this section, we have reviewed the scalar wave packet of the WDW equation for
the interior of the black hole, leading to
Yeom's interpretation. We have confirmed that the same interpretation is possible
for spherically symmetric and topological black holes in arbitrary dimensions.

\section{Spinor wave function inside a black hole}
\label{sec3}

In the previous work \cite{KAHS1,KAHS2}%
\footnote{A similar method has been developed by Robles-P\'erez \cite{Perez2}.}%
\footnote{The Eisenhart--Duval lift for quantum cosmology and quantum black
holes is also studied by Achour et al.\cite{ALOP}.},  the present authors
introduced the extended minisuperspace description with an additional degree of
freedom by a method of the Eisenhart--Duval lift
\cite{Eisenhart,Duval,DGH,CPC,Pettini,Cariglia,CGGH,FK,Finn,DSS}.
Using the covariance in the extended minisuperspace, we also
constructed an associated Dirac equation for spinor
wave function. The introduction of the
Dirac-type WDW equation (dubbed spinorial WDW equation) is originally motivated
for obtaining a positive-definite probability density
\cite{RS,Ryan,DHO,KO,SC,YH,HA,RAH,Moniz}. The Dirac equation in the
extended minisuperspace has a unique form and conformal covariance. 

In this section,
we apply the formulation to the present model for a black-hole interior.
In short, the realization of the plan is achieved by finding the metric of the
minisuperspace manifold such that the WDW equation is written by Laplace equation
($\nabla^2\Phi=0$), and drawing down the Dirac equation ($D\!\!\!\!/\,\Psi=0$) on
that manifold.
The Dirac-type equation naturally gives a positive defined probability density
($\Psi^\dagger\Psi$). The requirement of the covariance in the extended
minisuperspace resolves the problem of factor ordering and the ambiguity caused
by the Dirac square root
\cite{KAHS1,KAHS2}.

The essential idea of the Eisenhart--Duval lift is to introduce the metric
$G_{MN}$ of the extended space with an auxiliary dimension $Z$, that is, in the
present case,
\begin{equation}
G_{MN}dX^MdX^N={4r_s^{2(D-3)}e^{2Y}}(-dX^2+dY^2)+dZ^2\,,
\end{equation}
where $X^M=(X,Y,Z)$,%
\footnote{The geodesic equation on this metric gives the equation of
motion of the system under consideration 
\cite{Eisenhart,Duval,DGH,CPC,Pettini,Cariglia,CGGH,FK,Finn,DSS}.} 
and consider
the Laplace equation in the extended
minisuperspace:
\begin{equation}
\frac{1}{\sqrt{-G}}\partial_M(\sqrt{-G}G^{MN}\partial_N\Phi)=\left[
\frac{1}{4r_s^{2(D-3)}e^{2Y}}\left(-\frac{\partial^2}{\partial
X^2}+\frac{\partial^2}{\partial Y^2}
\right)+\frac{\partial^2}{\partial Z^2}\right]\Phi=0\,.
\end{equation}
Here, $G^{MN}$ is the inverse of $G_{MN}$, $G=-(4r_s^{2(D-3)})^2e^{4Y}$ is the
determinant of $G_{MN}$, and the derivatives are expressed as
$\partial_M\equiv\frac{\partial}{\partial X^M}$.
If we further impose an additional (first-order) constraint%
\footnote{The right-hand side of (\ref{2c}) can generally be $p^2\Phi$,
where $p$ is a constant, but we can choose $p=1$ without loss of generality
due to the constant scale invariance of the lifted metric or the arbitrary choice
of the time coodinate in the original spacetime \cite{KAHS1,KAHS2}}
\begin{equation}
\frac{\partial^2\Phi}{\partial Z^2}=-\Phi\,,
\label{2c}
\end{equation}
we can exactly reproduce the WDW equation (\ref{wdw}).

The WDW equation of Dirac-type for a spinor wave function $\Psi$ in the extended
minisuperspace is now naturally introduced by \cite{KAHS1,KAHS2}
\begin{equation}
D\!\!\!\!/\,\Psi\equiv\hat{\gamma}^MD_M\Psi\equiv\gamma^A e_A^MD_M\Psi=0\,,
\label{DE}
\end{equation}
where the constant gamma matrices in the flat spacetime $\gamma^A$ ($A=1,2,3$) are
$\gamma^1=\sigma^1$, $\gamma^2=i\sigma^2$, and $\gamma^3=i\sigma^3$
($\sigma^1$, $\sigma^2$, $\sigma^3$ denote the Pauli matrices). Note that
$\{\gamma^A,
\gamma^B\}=-2\eta^{AB}$, where $\eta^{AB}=\eta_{AB}=\mbox{diag.}(-1,1,1)$.
Here, the dreibein $e^A_M=\mbox{diag.} (2r_s^{D-3}e^{Y}, 2r_s^{D-3}e^{Y},
1)$ is defined through $\eta_{AB}e^A_Me^B_N=G_{MN}$,
and $e_A^M=\mbox{diag.} ((2r_s^{D-3})^{-1}e^{-Y}, (2r_s^{D-3})^{-1}e^{-Y},
1)$ is its inverse matrix. Subsequently, we find that $\{\hat{\gamma}^M,
\hat{\gamma}^N\}=-2G^{MN}$. The covariant derivative $D_M$ for the spin connection
$\omega_{MAB}$ is defined as
$D_M\equiv\partial_M+\frac{1}{4}
\omega_{MAB}\Sigma^{AB}$, where $\Sigma^{AB}\equiv-\frac{1}{2}[\gamma^A,
\gamma^B]$. The spin connection $\omega_{MAB}$ is given by
\begin{equation}
\omega_{MAB}=\frac{1}{2}e^N_A(\partial_Me_{NB}-\Gamma^L_{MN}e_{LB})
-(A\leftrightarrow B)\,,
\end{equation}
where the Christoffel symbol $\Gamma^L_{MN}$ is given by
\begin{equation}
\Gamma^L_{MN}=\frac{1}{2}G^{LP}(\partial_MG_{PN}+\partial_NG_{PM}
-\partial_PG_{MN})\,.
\end{equation}
In the present case, one can find $\omega_{X12}=-\omega_{X21}=-1$.

In the extended minisuperspace presently considered, we find that the Dirac
equation (\ref{DE}) is equivalent to
\begin{equation}
\left[\sigma^1\frac{\partial}{\partial X}+
i\sigma^2\left(\frac{\partial}{\partial Y}+\frac{1}{2}\right)+
i\sigma^3(2r_s^{D-3})e^{Y}\frac{\partial}{\partial Z}\right]\Psi=0\,.
\end{equation}
In order to reduce the equation to that of physical variables $X$ and $Y$,
we choose the additional constraint on $\Psi$:
\begin{equation}
\frac{1}{i}\frac{\partial}{\partial Z}\Psi=\Psi\,.
\end{equation}
Now, the Dirac equation reads in the matrix form, 
\begin{equation}
\left(
\begin{array}{cc}
-2r_s^{D-3}e^{Y} &
\frac{\partial}{\partial X}+\frac{1}{2}+\frac{\partial}{\partial Y}\\
\frac{\partial}{\partial X}-\frac{1}{2}-\frac{\partial}{\partial Y} &
2r_s^{D-3}e^{Y}
\end{array}
\right)\left(
\begin{array}{c}
\Psi_{+} \\ \Psi_{-}
\end{array}
\right)=\left(
\begin{array}{c}
0 \\ 0
\end{array}
\right)\,,
\label{DE1}
\end{equation}
where $\Psi={\Psi_{+}\choose\Psi_{-}}e^{iZ}$.
Note that the coordinate $Z$ is fictuous and has no effect on physical
quantities \cite{KAHS1,KAHS2}.%
\footnote{In fact, the probability density (\ref{pd}) is determined independently
of
$Z$.}
The fundamental solutions of the equation are found to be
\begin{equation}
\psi_{\pm,k}\propto\pm
K_{ik\pm\frac{1}{2}}(2r_s^{D-3}e^{Y})e^{-ikX}\,,
\end{equation}
and general solutions can be written as
\begin{equation}
\Psi_\pm(X,Y)=\int_{-\infty}^\infty f(k) \psi_{\pm,k}(X,Y)\,dk\,,
\label{sw}
\end{equation}
where $f(k)$ denotes the amplitude.

Since the inner product of two spinors in the curved minisuperspace can be defined
by
\begin{equation}
(\vartheta,\varphi)\equiv \int dY \sqrt{-G}|G^{XX}|^{1/2}
\vartheta^\dagger\varphi=2r_s^{D-3}\int dY\, e^Y\, 
\vartheta^\dagger\varphi
\,,
\end{equation}
the (positive-definite) probability density $\rho(X,Y)\propto e^Y\Psi^\dagger\Psi$
should takes the form%
\footnote{Note that the minisuperspace metric is Minkowskian, so we
can regard $X$ ($Y$) as a time (position) coordinate.
Therefore, we can describe the evolution of the probability measure $dP=\rho(X,Y)
dY$, where $dY$ is the `volume' element, in terms of the `time', $X$.} 
\begin{equation}
\rho(X,Y)\propto e^Y(|\Psi_+(X,Y)|^2+|\Psi_-(X,Y)|^2)\,.
\label{pd}
\end{equation}


Asymptotics of the modified Bessel function of the second
kind with complex order are known to be \cite{Tseng}
\begin{eqnarray}
K_{ik\pm\frac{1}{2}}(z)&\sim&\sqrt{2\pi}e^{-\frac{k\pi}{2}\pm
i\frac{\pi}{4}}(k^2-z^2)^{-\frac{1}{4}}\Biggl[\sqrt{\frac{k+z}{2z}}
\sin\left(\frac{\pi}{4}-\frac{1}{2}\sqrt{\nu^2-y^2}+k
\cosh^{-1}\frac{k}{z}\right)\nonumber \\
& &\quad\qquad\qquad\qquad\mp i\sqrt{\frac{k-z}{2z}}
\cos\left(\frac{\pi}{4}-\sqrt{k^2-z^2}+k
\cosh^{-1}\frac{k}{z}\right)\Biggr]\,,
\end{eqnarray}
as $k\rightarrow\infty$. This function has a similar oscillatory behavior
as $K_{ik}(z)$ (\ref{asym}).
The similarity in asymptotics of
$K_{ik}(z)$ and
$\sqrt{z}K_{ik\pm\frac{1}{2}}(z)$ also helps to construct wave packet solutions
for the Dirac-type WDW equation as well as the usual WDW equation.
The different behaviors of the functions $K_{ik}(z)$ and
$\sqrt{z}K_{ik\pm\frac{1}{2}}(z)$ seems to be especially revealed in the
vicinity of $X\approx 0$ in the form of the wave packets. Therefore, the
difference between wave-packet solutions of the usual WDW equation and the
Dirac-type WDW equation is expected to be found in the region of turning point of
the classical trajectory. 

\begin{figure}[ht]
\centering
\includegraphics
{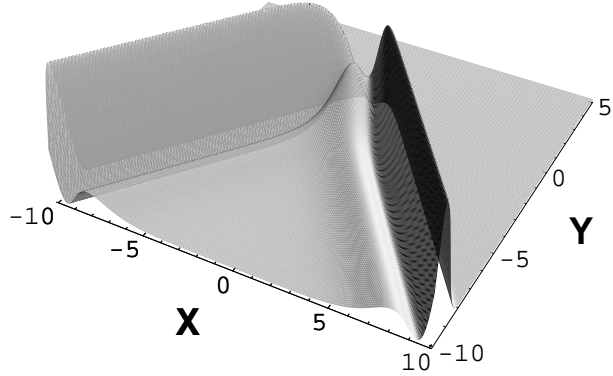}
\caption{The probability density $\rho_1(X,Y)$
for $r_s=1$.
(If you see jaggies in the plot, it is due to resolution limitations.)}
\label{fig2}
\end{figure}

We find that the simple choice $f(k)=ik$ in (\ref{sw}) leads to
\begin{eqnarray}
(\Psi_{1})_\pm(X,Y)&=&\pm\int_{-\infty}^\infty ik e^{-ikX}\,
K_{ik\pm\frac{1}{2}}(2r_s^{D-3}e^Y)\, dk\nonumber \\
&=&\pm\pi\Biggl[2r_s^{D-3}e^Y\sinh X\mp
\frac{1}{2}\Biggr]e^{\pm\frac{X}{2}}
e^{-2r_s^{D-3}e^Y\cosh X}\,.
\end{eqnarray}
The probability density $\rho_1(X,Y)=
e^Y(|(\Psi_1)_+(X,Y)|^2+|(\Psi_1)_-(X,Y)|^2)$
is exhibited in Fig.~\ref{fig2}.
The peak of the probability density reproduces the classical path where $|X|$ is
large, but a local hump of the density appears around $X=0$.
It would be appropriate to interpret this feature of the probability
density as ``Annihilation-to-Something'' process.

\begin{figure}[ht]
\centering
\includegraphics
{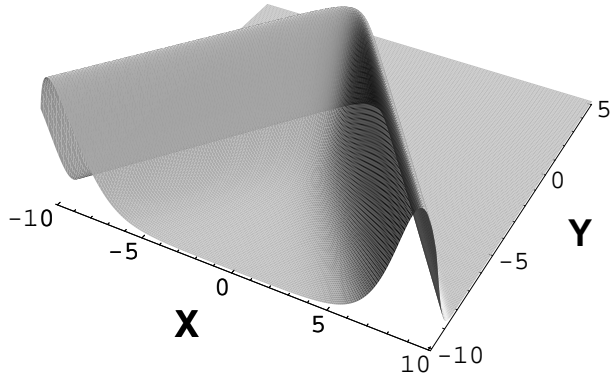}
\caption{The probability density $\rho_0(X,Y)$
for $r_s=1$.
(If you see jaggies in the plot, it is due to resolution limitations.)}
\label{fig3}
\end{figure}

On the other hand, another simple choice $f(k)=1$ in (\ref{sw}) leads to
\begin{eqnarray}
(\Psi_{0})_\pm(X,Y)&=&\pm\int_{-\infty}^\infty  e^{-ikX}\,
K_{ik\pm\frac{1}{2}}(2r_s^{D-3}e^Y)\, dk\nonumber \\
&=&\pm\pi e^{\pm\frac{X}{2}}
e^{-2r_s^{D-3}e^Y\cosh X}\,.
\end{eqnarray}
The probability density $\rho_0(X,Y)=
e^Y(|(\Psi_0)_+(X,Y)|^2+|(\Psi_0)_-(X,Y)|^2)$
is exhibited in Fig.~\ref{fig3}.
The peak of the density rather exactly reproduces the classical path, even
where $X\approx 0$. 
It is said that there is ``no annihilation'' process, since the wave packet
represents a classical trajectory without interruption.

Although only simple amplitude functions have been assumed here,
it turns out that the behavior of the density function in the region where the
quantum effect is expected ($X\approx 0$) has more diverse for spinor waves than
for scalar waves.

In this section, we have introduced the spinorial WDW equation for the interior
of the black hole and constructed the typical wave packets. We have found that, in
our spinorial formulation, we can create wave packets with various
interpretations.
This means that, even if we assume wave packets which correspond to classical
trajectories in the asymptotic region and limit ourselves to simple amplitude
functions,  the ``Annihilation-to-Nothing'' process (proposed by Yeom et al.) is
not unique in the formulation of spinorial wave functions.

\section{Summary and discussion}
\label{conclusion}

We have solved the spinorial wave function (as well as the scalar wave
function as a comparison) of the interior of the Schwarzschild--Tangherlini
and planar topological black holes in $D$-dimensional spacetime,
 by using the extended minisuperspace.
We have constructed the wave packets of the spinor wave function,
which correspond to the classical trajectories in the asymptotic region near the
horizon and the singularity.

The behavior of the scalar wave packet near $X\approx 0$ suggests
the ``Annihilation-to-nothing'' interpretation proposed by Yeom et
al.~\cite{BBCCY,Yeom1,Yeom2,BCY}. The probability density of the spinor wave
function is slightly different from that of the scalar wave function, in the
vicinity of $X=0$ for typical examples of the packets. Thus, we 
speculate that alternative interpretation of ``no annihilation'' or
``annihilation-to-something'' is possible. 

Since our spinorial formulation leads to a positive definite probability
density, it will be applied to further analysis of the expectation
values \cite{Hartnoll} near the singularity (even without using wave packets). We
shall leave the issue for future study.

We should also incorporate possible noncommutativity in the variables of the
minisuperspace (see, for example, Ref.~\cite{GOR} and references therein). The
approach by considering loop quantum cosmology (see
Refs.~\cite{Modesto,AB,CS,SG,GLS,OZZWW} and references therein) (or polymer
quantization) may be effective for the analysis at the singularity and the
pursuit of the possibility of singularity resolution from limiting curvature
deformations. We hope that the study in the aforementioned directions will be
advanced in future.

\bibliographystyle{apsrev4-1}


\end{document}